\documentclass[aps,prl,twocolumn,superscriptaddress,showpacs,amsfonts,amsmath,amssymb]{revtex4}
\usepackage{graphicx}
\usepackage{dcolumn}
\usepackage{bm}

\allowdisplaybreaks[1]

\begin{document}
\title{Nonlocal stabilization of nonlinear beams in a self-focusing atomic vapor }

\author{S. Skupin}
\altaffiliation{Present address: D\'epartement de Physique Th\'eorique et Appliqu\'ee,
CEA/DIF, B.P. 12, 91680 Bruy\`eres-le-Ch\^atel, France}
\email{stefan.skupin@cea.fr}
\affiliation{Research School of Physical Sciences and Engineering,
Australian National University, Canberra, ACT 0200, Australia}

\author{M. Saffman}
\affiliation{Department of Physics, University of Wisconsin, 1150 University Avenue,
Madison, WI 53706}

\author{W. Kr\'olikowski}
\affiliation{Research School of Physical Sciences and Engineering,
Australian National University, Canberra, ACT 0200, Australia}

\date{\today}

\begin{abstract}
We show that ballistic transport of optically excited atoms in an atomic vapor provides a nonlocal nonlinearity 
which stabilizes the propagation of vortex beams and
higher order modes in the presence of a self-focusing nonlinearity. Numerical experiments demonstrate stable 
propagation of lowest and higher order vortices over
a hundred diffraction lengths, before dissipation leads to decay of these structures.
\end{abstract}

\maketitle


The propagation and dynamics of localized nonlinear waves is a  subject of great interest in a range of physical settings 
stretching from nonlinear optics to plasmas and ultracold atomic gases~\cite{Nail_book,BEC_soliton}. The structure and stability of nonlinear optical modes  is determined by the interplay of the radiation field with the functional form of the material nonlinearity~\cite{KivsharAgrawal}. 
In the case of optical beams the nonlinear response can be described in terms of the induced change in the refractive index $n$ which is often approximated as a local function of the wave intensity, i.e. $n({\bf r})= n(I({\bf r})).$ However, in many real physical systems the nonlinear response is spatially nonlocal which means that the refractive index depends on the beam intensity in the neighborhood of each spatial point. 
This can be phenomenologically expressed as  
 $n({\bf r})= \int d{\bf r'}\, K({\bf r},{\bf r'}) I({\bf r'})$, where the response kernel $K$
depends on the particular model of nonlocality~\cite{snyder97}. 
 
It has been shown that nonlocality drastically
affects the stationary structure and dynamics of spatial solitons,
leading to such effects as collapse arrest of high intensity beams
and stabilization of otherwise unstable complex solitonic
structures~\cite{Kolchugina,krolikowski98,JOB,skupin06}.
Nonlocality is often the consequence of transport processes which include 
  atom or  heat diffusion in atomic
vapors~\cite{nonlocalvapor}, plasma~\cite{Litvak75} and thermal
media~\cite{Dabby68}, or charge drift in photorefractive
crystals~\cite{zozulya94}.  In addition  long range interactions are responsible for 
a nonlocal response in  liquid crystals~\cite{Conti03} or dipolar
Bose Einstein condensates~\cite{Santos00}. 

Hot atomic vapors are an important and widely
used nonlinear medium. The nonlocal nonlinear  response of atomic vapors has previously only been 
 associated with state dependent transport of ground state atoms which possess a multilevel structure~\cite{nonlocalvapor}. In
this letter we introduce a new and significant mechanism of nonlocality  in atomic vapors which is provided by the   ballistic transport of excited atoms and is important even for the simplest case of an idealized two-level atom. We show using parameters representative of beam propagation in Rubidium vapor that ballistic transport 
 plays a dramatic role leading to 
stabilization of otherwise unstable vortex modes in the presence of a self-focusing nonlinearity.

Prior to introducing a model for the nonlocal character of the refractive index we first recall the main features of beam propagation in a hot atomic vapor. 
We consider a scalar traveling wave $E=\frac{{\mathcal E}(x,y,z)}{2}e^{\imath(kz-\omega t)}+c.c.$ For all parameters of interest the refractive index is $n\simeq 1$ so the wave intensity is 
$I\simeq \frac{\epsilon_0 c}{2}|{\mathcal E}|^2.$
In the slowly 
varying envelope approximation the paraxial wave  equation is 
\begin{equation}
\frac{\partial {\mathcal E}}{\partial z}-\frac{i}{2k}\nabla_\perp^2 {\mathcal E} =
-\frac{k\chi_0''}{2} {\mathcal E}
+i\frac{k}{2}\left[\chi_{\rm nl}'(I)+i \chi_{\rm nl}''(I)\right] {\mathcal E},
\label{eq.pwe2}
\end{equation}
where $k=\omega/c.$ 
The susceptibilities $\chi'$ and $\chi''$ depend on atomic parameters. 
We assume a two-level atomic model for which the   scattering cross section   is
$\sigma = (3\lambda_a^2/2\pi)\left(1+4\Delta_0^2/\gamma^2+I/I_s\right)^{-1},$
and the index of refraction is $n=1-n_a (\sigma/k) (\Delta_0/ \gamma),$ 
where $\lambda_a$ is the transition wavelength, $n_a$ is the atomic density, $\gamma$ is the full width at half maximum (FWHM) natural linewidth,  $\Delta_0=\omega-\omega_{a}$
is the detuning  between the optical frequency $\omega$ and the atomic transition frequency $\omega_a= 2\pi c/\lambda_a$, 
and $I_s$ is the saturation intensity. For a probe beam propagating along $\hat z$ in a hot vapor we make the 
replacement  $\Delta_0\rightarrow \Delta=\Delta_0-kv_z.$ Averaging over a Maxwell-Boltzmann velocity distribution 
at temperature $T$ gives an expression for the complex susceptibility which can be separated into a constant part and a part which depends on intensity. The results can be written in the form~\cite{close67} $\chi_0'' = \chi_0 {\rm Im}[Z(a+ib)]$,
\begin{subequations}
\label{eq.pwe3}
\begin{align}
\chi_{\rm nl}'(I) & = \chi_0\left\{{\rm Re}[Z(a+ib_I)]-{\rm Re}[Z(a+ib)]\right\},
\label{eq.pwe2a}
\\
\chi_{\rm nl}''(I) & = \chi_0\left\{ \frac{{\rm Im}[Z(a+ib_I)] }{\sqrt{1+I/I_s}}
-{\rm Im}[Z(a+ib)] \right\},
\label{eq.pwe2b}
\end{align}
\end{subequations}
where $\chi_0=n_a  6\pi b \, c^3/\omega_a^3$, 
$a=2\sqrt{\ln2}\Delta_0/\omega_D$, 
$b= \sqrt{\ln2}\gamma/\omega_D,$ $b_I=b\sqrt{1+I/I_s}$,
and
$\omega_D=k\sqrt{8 \ln2 k_B T/m}$  is the FWHM of the Doppler profile 
for an atom of mass $m$. 
The plasma dispersion function is given by   
$ Z(z)=i\sqrt\pi e^{-z^2}{\rm Erfc}(-iz)
$ where $\rm Erfc(z)=1-(2/\sqrt\pi)\int_0^z dt\,e^{-t^2}$ is the complementary error function.

\begin{figure}[!t]
\includegraphics[width=1.\columnwidth]{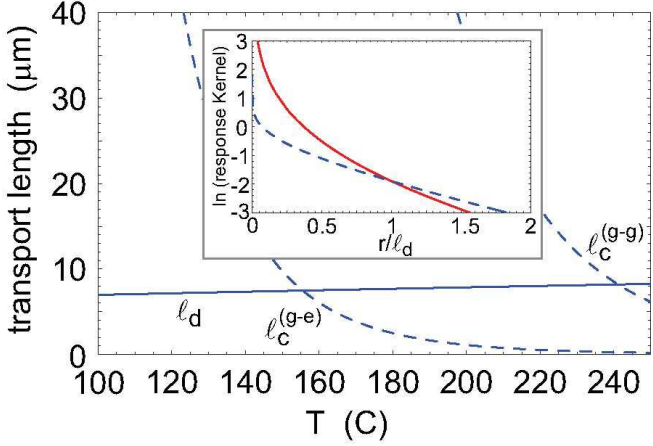}
\vspace{-.3cm}
\caption{(color online) Characteristic length scales  $\ell_c^{(g-g)}$, $\ell_c^{(g-e)},$ and $\ell_d$ in a Rb vapor cell as a function of temperature. The inset shows the logarithm of the ballistic response function $\sim (\ell_d/r)
\int_{0}^\infty d\xi\,\ e^{-r/(\ell_d\xi)}e^{-\xi^2}$ (solid blue line) and
the response function $K_0(r/\ell_d)$ for a 2D diffusive nonlocal equation~\cite{skupin06}  (dashed red line) as a function of 
$r/\ell_d.$
The response functions have been scaled to be equal at $r=\ell_d.$  }
\label{fig.rbtransport}
\end{figure}

For broad optical beams numerical solutions of Eqs.\ (\ref{eq.pwe2}) and (\ref{eq.pwe3})  give an  accurate description of  
 propagation effects in an atomic vapor. 
The physical effect leading to the nonlinear optical response is the transfer of population from the ground to the excited 
state and the creation of coherence between these states. Although motional effects are  accounted for as regards the Doppler smearing of the transition frequency, 
atomic motion also results in transport of excited atoms which leads to a nonlocal response. 
The degree of nonlocality depends on the characteristic length scales associated 
with the transport of excited state atoms. The first length scale is the  mean free path atoms travel before a Rb-Rb collision occurs. 
This is given by  $\ell_c=(\sqrt2 n_a \sigma)^{-1}$, where $\sigma$ is the collisional cross section.
For collisions of  ground state Rb atoms we use~\cite{thomas99} 
$\sigma_{\rm g-g}=2.5\times 10^{-17}~\rm m^2$. The cross section for collisions 
between excited and ground state collisions is much larger since these collisions occur via a long range dipole-dipole  interaction~\cite{lewisreview}. The energy averaged cross section  is 
$\sigma_{g-e}\sim \frac{1.8\times 10^{-14}}{\sqrt T}~\rm  K^{1/2} m^2$. 
 The second length scale is  $\ell_d=\tilde v\tau$, which is 
 the distance traveled by an atom moving at the most probable  speed
 $\tilde v= \sqrt{2 k_B T/m}$  in the $1/e$ lifetime $\tau$ of the excited state. For the $^{87}$Rb 5P$_{3/2}$ level $\tau \simeq 26~\rm ns$.
 Figure \ref{fig.rbtransport} shows that for $T<155^{\circ}\rm C$  
 the ballistic transport length for excited atoms is  $\ell_d\sim 7.5~\mu\rm m$ and $\ell_d<\ell_{c}^{(g-e)},\ell_{c}^{(g-g)}$.
Thus for these temperatures 
transport of excited atoms is ballistic with a length scale of $\ell_d.$ We note that the density at
$T=155^{\circ}\rm C$ is  $n_a= 10^{20}~\rm m^{-3}$ which is several orders of magnitude smaller than densities for which nonlocal effects
due to the Lorentz local field are important\cite{boydmalcuit}.

We wish to find an expression for the nonlocal material response that depends on the parameters $\tau$ and $\tilde v.$ 
We write the total atomic density $n_a=n_g+n_e$ as the sum of ground and excited state partial densities and introduce  rate equations of the form
\begin{subequations}
\begin{align}
\frac{\partial n_g}{\partial t}&=-\frac{ I\alpha(I)}{\hbar \omega}+\gamma n_e + {\mathcal L}_g[n_g]
\\
\frac{\partial n_e}{\partial t}&=\frac{ I\alpha(I)}{\hbar \omega}-\gamma n_e +  {\mathcal L}_e[n_e],
\end{align}
\end{subequations}
where ${\mathcal L}_g, {\mathcal L}_e$ are, as yet unknown, linear operators for  ground and excited state atoms and $\alpha(I)=k[\chi_0''+\chi_{\rm nl}''(I)]$ is the  absorption coefficient.
If we assume that the total 
density is unchanged by the presence of the laser field (this is a reasonable assumption in hot vapors, but not for cold atoms) we must 
have ${\mathcal L}_g[n_g] = - {\mathcal L}_e[n_e]$.
If excited state transport were a diffusive process we would 
have ${\mathcal L}_e[n_e]=D_e\nabla^2 n_e$ and on dimensional grounds 
$D_e\sim (\bar v\tau)^2/\tau\sim \ell_d^2/\tau.$

The situation in the ballistic regime is different. The collisionless Boltzmann equation for the density of excited atoms 
is $\frac{d n_e}{dt}=\frac{\partial n_e}{\partial t}+ {\bf v}\cdot\nabla_{\bf r}n_e.$
Working within the paraxial approximation we are interested in the two-dimensional problem where ${\bf r}=x\hat x + y\hat y$ 
and ${\bf v}=v_x\hat x + v_y\hat y.$
The Green function is found by solving 
\begin{equation}
\frac{\partial n_e}{\partial t}+ {\bf v}\cdot\nabla_{{\bf r}}n_e=\delta(t)n_0({\bf r},{\bf v}).
\label{eq.cb2}
\end{equation}
The time dependent solution of (\ref{eq.cb2}) is 
$ n_e({\bf r},{\bf v},t)=H(t)n_0({\bf r}-{\bf v} t,{\bf v})$ where $dH/dt=\delta(t).$
Now consider the solution for  $n_0({\bf r},{\bf v})=\delta({\bf r}_0)f({\bf v})$, with $f({\bf v})=(m/2\pi k_B T)e^{- v^2/\tilde v^2}$ the 
two-dimensional thermal velocity distribution.  The solution  is 
\begin{equation}
G({\bf r},t;{\bf r}_0,t_0)= \delta({\bf r}-{\bf r}_0-{\bf v} (t-t_0))\frac{m e^{- v^2/\tilde v^2}.}{2\pi k_B T}.
\end{equation}
The Green function for the spatial response is then 
\begin{align}
G_r({\bf r},t;{\bf r}_0,t_0)&= \int d{\bf v}\, \delta({\bf r}-{\bf r}_0-{\bf v} (t-t_0))\frac{me^{-v^2/\tilde v^2}}{2\pi k_B T}\nonumber\\
&=\frac{ 1}{\pi \tilde v^2}\frac{1}{(t-t_0)^2}e^{- |{\bf r}-{\bf r}_0|^2/(\tilde v^2 (t-t_0)^2)}.
\end{align}
This calculation neglects the fact that the excitation decays with rate $\gamma$. We can account for this by including a factor 
of $e^{-\gamma t}$ in the Green function to arrive at 
\begin{equation}
\label{greenr}
G_r({\bf r},t;{\bf r}_0,t_0;\gamma)=
\frac{ 1}{\pi \gamma^2 \ell_d^2}\frac{e^{-\gamma(t-t_0)}}{(t-t_0)^2}e^{- |{\bf r}-{\bf r}_0|^2/(\tilde v^2(t-t_0)^2)}.
\end{equation}
The  Green function is parameterized by the decay rate $\gamma.$
Since the rate of local excitation is $\alpha I /\hbar\omega$ the   spatial distribution of $n_e$ is  given by
\begin{equation}
\begin{split}
n_e({\bf r},t)&= 
\int_{-\infty}^t dt_0  \int d{\bf r}_0\, G_r({\bf r},t;{\bf r}_0,t_0;\gamma) \\
&\quad\times
\frac{\alpha[I({\bf r}_0,t_0)]}{\hbar\omega}I({\bf r}_0,t_0).
\end{split}
\end{equation}
The  response to a temporally constant field which is a delta function in space, $I=I_0\delta({\bf r}_0)$, is, with Eq.\ (\ref{greenr}),
\begin{equation}
n_e({\bf r})=\frac{\alpha(I_0)I_0}{\hbar\omega}
\frac{1}{\pi\tilde v r }
\int_{0}^\infty d\xi\,\ e^{-\gamma r/(\tilde v\xi)}e^{-\xi^2}.
\end{equation}
The Green function for the steady state  spatial response which has units of $\rm s/m^2$ is thus 
\begin{equation}
G({\bf r},{\bf r}_0;\gamma)=
\frac{1}{\pi\tilde v r }
\int_{0}^\infty d\xi\,\ e^{-\gamma r/(\tilde v\xi)}e^{-\xi^2},
\label{eq.greenss}
\end{equation}
where $r=|{\bf r}-{\bf r}_0|.$ 
The result is plotted in the inset of Fig.\ \ref{fig.rbtransport} as a function of the scaled 
coordinate $r/\ell_d.$ We see, not unexpectedly, that the ballistic response falls off much more rapidly than the diffusive response. 
Note that $\int d{\bf r}\,  G({\bf r},0,\gamma)=1/\gamma,$ since the time integrated response exponentially weights the 
input over a time window $\tau=1/\gamma.$
The spatial  Fourier transform of the Green function is given by
${\mathcal F}[G]=
\frac{\sqrt\pi}{\gamma }
\frac{e^{1/(k \ell_d )^2 }}{k \ell_d}
{\rm Erfc}[1/(k \ell_d )]$
which  is well behaved with  $\lim_{k\rightarrow 0}{\mathcal F}[G]=1/\gamma.$

To complete the theoretical formulation of the wave propagation problem we need to calculate the nonlocal structure of the 
susceptibility $\chi_{\rm nl}$. 
The imaginary part of the nonlinear  susceptibility is proportional to the differential density of excited and ground state 
atoms which decays with rate $\gamma.$ When the intensity is uniform in space 
the susceptibility satisfies the relaxation equation 
$d\chi_{\rm nl}''/dt = - \gamma (\chi_{\rm nl}'' - \overline{\chi_{\rm nl}''})$
where the overbar denotes the steady state value of the susceptibility.
When $I$ is spatially varying we can use the Green function to write the stationary response as 
\begin{multline}
\chi_{\rm nl}''({\bf r})
=\gamma\chi_0 
\int d{\bf r}_0\, G({\bf r},{\bf r}_0;\gamma)
\\
\times\left\{ \frac{{\rm Im}[Z(a+ib_I[I({\bf r}_0)])] }{\sqrt{1+I({\bf r}_0)/I_s}}
 -{\rm Im}[Z(a+ib)] 
\right\}.
\label{eq.chiimagnonlocal}
\end{multline}

The real part of the susceptibility is proportional to the coherence between ground and excited states which decays with rate $\gamma/2.$ 
The Green function to be used for $\chi_{\rm nl}'$ is thus
$G({\bf r},{\bf r}_0,\gamma/2)$ and we can write
\begin{multline}
\chi_{\rm nl}'({\bf r})=\frac{\gamma\chi_0}{2} 
\int d{\bf r}_0\, G({\bf r},{\bf r}_0;\gamma/2)
\\
\times\left\{{\rm Re}[Z(a+ib_I[I({\bf r}_0)])]   -
{\rm Re}[Z(a+ib)]
\right\}.
\label{eq.chirealnonlocal}
\end{multline}
Equations (\ref{eq.chiimagnonlocal},
\ref{eq.chirealnonlocal}) together with the Green function  Eq.\ (\ref{eq.greenss}) and  the wave equation 
(\ref{eq.pwe2}) are the main theoretical result of this paper. They 
 constitute a full description of 
time-independent wave propagation in a two-level atomic vapor including Doppler broadening and transport induced nonlocality. 

The question of whether or not the ballistic transport is sufficient to stabilize nonlinear modes can be investigated by beam propagation calculations. We use parameters corresponding to off-resonant propagation in a high temperature Rb cell ($\lambda=780~\rm nm$,
$\gamma/2\pi  = 6.07~\rm MHz$,  $\Delta_0/2\pi = 1.46 ~\rm GHz$, 
$I_s=16.7~\rm W/m^2$, $T\simeq155^{\circ}\rm C$, $n_a=10^{20}~{\rm m}^{-3}$, $\ell_d=7.5~\mu\rm m$), which result in the dimensionless 
parameters $a=4$, $b=0.0083$,  and $\chi_0=0.03$. We used this set throughout all the simulations presented in this paper. In the conservative system ($\chi_0''=\chi_{nl}''=0$) all modes we tried  (ground state, single charged vortex, dipole, double charged vortex) turned out to be stable  if the power $P = \int  I({\bf r})d{\bf r}, $ is high enough. At least for the latter two modes this is quite remarkable, since they are known to be  
unstable (or only stable in a small power window) for other nonlocal models~\cite{Yakimenko05,skupin06,Lashkin06}. We attribute this enhanced stabilization to the combination of {\em nonlocality and nonlinear saturation}. In fact, we inserted 
an artificial nonlinear saturation in the nonlocal thermal model used in Ref.~\cite{Yakimenko05} and found that the double charged vortex becomes stable as well. However, it is worth pointing out that {\em nonlinear saturation without nonlocality} does not stabilize higher order nonlinear modes~\cite{Bigelow04}. The local Eqs.\ (\ref{eq.pwe2},\ref{eq.pwe3}) feature a stable ground state only.

\begin{figure}[!t]
\includegraphics[width=1.\columnwidth]{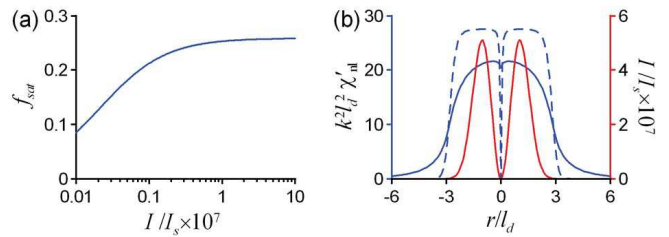}
\caption{\label{fig.chinl}(color online)  (a) Saturation function $f_{sat}={\rm Re}[Z(a+ib_I)]-{\rm Re}[Z(a+ib)]$ for $a=4$ and $b=0.0083$. (b) Nonlocal single charged vortex mode with power $P\simeq0.4~\rm W$ (red line and red axis). The solid blue line shows the {\em nonlocal} nonlinear index $\chi_{\rm nl}'$ computed from Eq.\ (\ref{eq.chirealnonlocal}), the dashed blue line the {\em local} one computed from Eq.\ (\ref{eq.pwe2a}) (blue axis).}
\end{figure}

Figure \ref{fig.chinl} illustrates both saturation and nonlocality for the nonlocal single charged vortex mode. If we consider only the saturation effect shown in Fig.\ \ref{fig.chinl}(a) the resulting nonlinear index is the dashed blue line in Fig.\ \ref{fig.chinl}(b). Together with the nonlocal kernel $G$ [red line in Fig.\ \ref{fig.rbtransport} inset] we get the solid blue line, showing some filling in of the central dip in the index profile, and the formation of a broader ``nonlocal waveguide".

\begin{figure}[!t]
\includegraphics[width=.95\columnwidth]{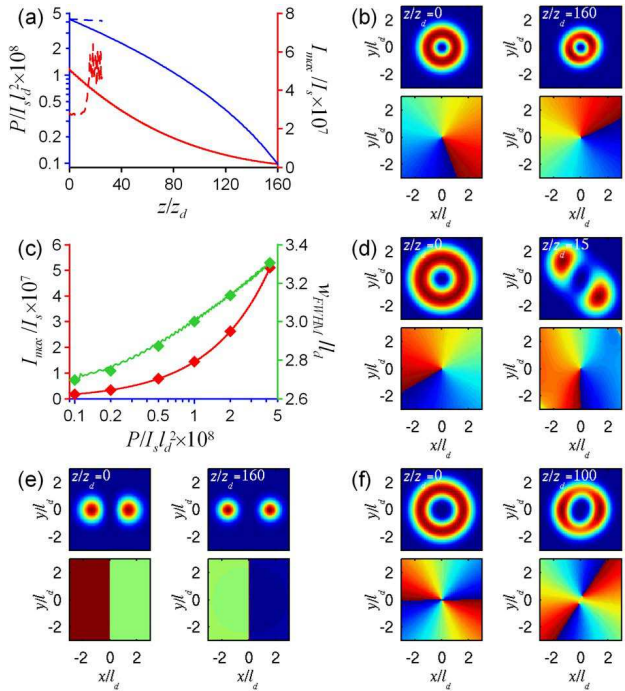}
\caption{\label{fig.numerics}(color online)  (a) {\em Nonlocal} (solid lines) and {\em local} (dashed lines) dissipative propagation of the single charged vortex mode with input power $0.4~\rm W$. The blue lines and blue axis show the beam power, the red lines and red axis the maximal intensity versus propagation distance.
(b) Intensity and phase distribution of the {\em nonlocal} single charged vortex at input $z=0$ and at $z=160z_d$ just before it decays.
(c) Maximal intensity (red) and FWHM (green) of the {\em nonlocal} single charged vortex as a function of beam power. The solid lines are computed upon propagation, the diamonds from stationary numerical solutions of the conservative problem. 
(d) Intensity and phase distribution of the {\em local} single charged vortex at $z=0$ and at $z=15z_d$ when it decays.
(e) Intensity and phase distribution of the {\em nonlocal} dipole mode at $z=0$ and at $z=160z_d$ just before it decays.
(f) Same for the {\em nonlocal} double charged vortex at $z=0$ and at $z=100z_d$.}
\end{figure}

The limiting mechanism with respect to long distance propagation of higher order nonlocal nonlinear modes is not destabilization but dissipation. The action of both $\chi_0''$ and $\chi_{nl}''$ is not negligible over one diffraction length $z_d=2k\ell_d^2$~\footnote{Since typical transverse length scales of the nonlinear modes under consideration are of the order $\ell_d$, this definition of the diffraction length makes sense.}.
As an illustrative example, the propagation of the {\em nonlocal} single charged vortex mode, is shown in Fig.\ \ref{fig.numerics}(a). As input power we use about $0.4~\rm W$. Note the clearly visible influence of the nonlinear term $\chi_{nl}$ in the blue power curve. The nonlocal vortex survives a propagation distance of more than $150z_d$ [see Fig.\ \ref{fig.numerics}(b)]. 
For comparison, the propagation of the {\em local} vortex with the same input power is shown in dashed lines in Fig.\ \ref{fig.numerics}(a).  This  vortex  disintegrates after  less than $15z_d$ [see Fig.\ \ref{fig.numerics}(d)]. Hence, we clearly see that the stabilization is due to nonlocality. With the same input power of about $0.4~\rm W$, we also observed a robust nonlocal dipole [see Fig.\ \ref{fig.numerics}(e)] and double charged vortex [see Fig.\ \ref{fig.numerics}(f)].

The key feature enabling robust nonlocal dissipative propagation over a hundred diffraction lengths is the above mentioned stability for high powers. Starting in the stable power regime, dissipation makes the nonlinear mode ``glide down" the family branch until it reaches powers in the unstable regime. Figure \ref{fig.numerics}(c) confirms this property by comparing maximal intensity and FWHM obtained upon propagation with values found from exact numerical solution  of the conservative problem using the method described in~\cite{skupin06}. The solid lines are obtained upon propagation, which explains the small oscillations in the curves.

In conclusion, we have shown that ballistic transport of optically excited atoms in a thermal vapor provides a generic nonlocal nonlinearity which can stabilize the propagation of vortices and other 
higher order modes in a self-focusing medium. For sufficiently high power we found a stable dipole mode and single and double charged vortices. In realistic models dissipation is not negligible. Nevertheless, numerical experiments demonstrate robust propagation over a hundred or more diffraction lengths. This is possible due to adiabatic conversion into solitons with lower power, but of the same family.

Numerical simulations were performed on the SGI Altix 3700 Bx2 cluster of the Australian Partnership for Advanced Computing
and on the IBM p690 cluster (JUMP) of the Forschungs-Zentrum in J\"ulich, Germany.


\end{document}